\documentclass[sn-nature]{sn-jnl}

\usepackage{graphicx}%
\usepackage{multirow}%
\usepackage{amsmath,amssymb,amsfonts}%
\usepackage{amsthm}%
\usepackage{mathrsfs}%
\usepackage[title]{appendix}%
\usepackage{xcolor}%
\usepackage{textcomp}%
\usepackage{manyfoot}%
\usepackage{booktabs}%
\usepackage{algorithm}%
\usepackage{algorithmicx}%
\usepackage{algpseudocode}%
\usepackage{listings}%
\usepackage{url}%
\usepackage[normalem]{ulem}%
\theoremstyle{thmstyleone}%
\theoremstyle{thmstyletwo}%

\theoremstyle{thmstylethree}%

\makeatletter
\def\@oddhead{Published in Scientific Reports 13,16091(2023) \url{https://www.nature.com/articles/s41598-023-42164-4}}
\makeatother

\begin{document}

\title[Observing Cosmic-Ray Extensive Air Showers with a Silicon Imaging Detector]{Observing Cosmic-Ray Extensive Air Showers with a Silicon Imaging Detector}

\author*[1]{\fnm{Satoshi} \sur{Kawanomoto}}\email{kawanomoto.satoshi@nao.ac.jp}

\author*[1]{\fnm{Michitaro} \sur{Koike}}\email{michitaro.koike@nao.ac.jp}

\author[2]{\fnm{Fraser} \sur{Bradfield}}

\author*[2,3]{\fnm{Toshihiro} \sur{Fujii}}\email{toshi@omu.ac.jp}

\author[4]{\fnm{Yutaka} \sur{Komiyama}}

\author[5,6]{\fnm{Satoshi} \sur{Miyazaki}}

\author[7]{\fnm{Tomoki} \sur{Morokuma}}

\author[8,9,10]{\fnm{Hitoshi} \sur{Murayama}}

\author[11,12]{\fnm{Masamune} \sur{Oguri}}

\author[5]{\fnm{Tsuyoshi} \sur{Terai}}

\affil[1]{\orgname{National Astronomical Observatory of Japan}, \orgaddress{\street{Osawa}, \city{Mitaka}, \state{Tokyo}, \postcode{181-8588}, \country{Japan}}}

\affil[2]{\orgdiv{Graduate School of Science}, \orgname{Osaka Metropolitan University}, \orgaddress{\street{Sumiyoshi}, \state{Osaka}, \postcode{558-8585}, \country{Japan}}}

\affil[3]{\orgdiv{Nambu Yoichiro Institute of Theoretical and Experimental Physics}, \orgname{Osaka Metropolitan University}, \orgaddress{\street{Sumiyoshi},  \state{Osaka}, \postcode{558-8585}, \country{Japan}}}

\affil[4]{\orgname{Department of Advanced Sciences, Faculty of Science and Engineering,
Hosei University}, \orgaddress{\street{Kajino-cho}, \city{Koganei}, \state{Tokyo}, \postcode{184-8584}, \country{Japan}}}

\affil[5]{\orgname{Subaru Telescope, National Astronomical Observatory of Japan}, \orgaddress{\city{Hilo}, \state{Hawai`i}, \postcode{96720}, \country{USA}}}

\affil[6]{\orgname{SOKENDAI, Graduate University for Advanced Studies}, \orgaddress{\street{Osawa}, \city{Mitaka}, \state{Tokyo}, \postcode{181-8588}, \country{Japan}}}

\affil[7]{\orgname{Planetary Exploration Research Center, Chiba Institute of Technology}, \orgaddress{\street{Tsudanuma}, \city{Narashino}, \state{Chiba}, \postcode{275-0016}, \country{Japan}}}

\affil[8]{\orgname{Berkeley Center for Theoretical Physics, University of California}, \orgaddress{\city{Berkeley}, \state{California}, \postcode{94720}, \country{USA}}}

\affil[9]{\orgname{Theory Group, Lawrence Berkeley National Laboratory}, \orgaddress{\city{Berkeley}, \state{California}, \postcode{94720}, \country{USA}}}

\affil[10]{\orgname{Kavli Institute for the Physics and Mathematics of the Universe (WPI), University of Tokyo}, \orgaddress{\city{Kashiwa}, \state{Chiba}, \postcode{277-8583}, \country{Japan}}}

\affil[11]{\orgname{Center for Frontier Science, Chiba University}, \orgaddress{\city{Inage}, \state{Chiba}, \postcode{263-8522}, \country{Japan}}}

\affil[12]{\orgname{Department of Physics, Graduate School of Science, Chiba University}, \orgaddress{\city{Inage}, \state{Chiba}, \postcode{263-8522}, \country{Japan}}}

\abstract{
Extensive air showers induced from high-energy cosmic rays provide a window
into understanding the most energetic phenomena in the universe. We \textcolor{black}{present} a new method for observing these showers using the silicon imaging detector
Subaru Hyper Suprime-Cam (HSC). This method has the advantage of being able
to measure individual secondary particles. \textcolor{black}{When paired with a surface detector array, silicon imaging detectors like Subaru HSC will be useful for studying the
properties of extensive air showers in detail.} The following report outlines the
first results of observing extensive air showers with Subaru HSC. The potential
for reconstructing the incident direction of primary cosmic rays is demonstrated
and possible interdisciplinary applications are discussed.\\\\\\
\textbf{Journal:} Published in Scientific Reports 13,16091(2023)\\
\textbf{URL:} \url{https://www.nature.com/articles/s41598-023-42164-4}\\
\textbf{DOI:} 10.1038/s41598-023-42164-4
}


\maketitle

\section{Cosmic-ray extensive air showers and surface detector arrays}
Cosmic rays are energetic particles from the universe discovered by V.F. Hess in 1912 \cite{Hess:1912srp}. Their origin, acceleration mechanisms and mass composition at the highest energies are still largely unknown. In particular, determining the origin of high-energy cosmic rays is considered to be among the most important problems in modern astrophysics~\cite{AlvesBatista:2019tlv,Coleman:2022abf}. The flux of cosmic rays arriving at Earth follows a power-law function of roughly $E^{-3}$, where $E$ is the primary energy of the cosmic rays. As such, high-energy cosmic rays are extremely rare and are typically studied indirectly. This is accomplished by measuring the showers of secondary particles produced when primary cosmic rays interact with atmospheric nuclei, called ``extensive air showers'' (EASs).

\textcolor{black}{Several methods are used to observe EASs, such as the detection of fluorescence light~\cite{Tokuno:2012mi,PierreAuger:2015eyc} and radio emission~\cite{PierreAuger:2016vya,Buitink:2016nkf,Schroder:2016hrv} from the electromagnetic component of an EAS. Another method} is the use of surface detector \textcolor{black}{(SD)} arrays - collections of ground based particle detectors spread across several hundreds or thousands of square kilometers~\cite{PierreAuger:2015eyc,bib:tasd,bib:tax4_nim}. Such arrays measure the arrival times and densities of the secondary particles from an EAS at ground level. This information is used to reconstruct \textcolor{black}{the} arrival direction, primary energy \textcolor{black}{and mass composition information} of the initial cosmic ray. As an example, the Telescope Array experiment in Utah, USA, uses 507 plastic scintillators, deployed with a spacing of 1.2\,km, to form a detecting area of $\sim$700\,km$^2$ to cosmic rays with energies above 10$^{19}$\,eV~\cite{bib:tasd}. By combining measurements from several detectors, the energy and arrival direction of the primary cosmic ray are determined. However, as is typical for surface detectors, these scintillators are unable to measure the properties of individual secondary particles during normal operation. Such information would be useful for identifying the primary particle and studying high-energy particle interactions\textcolor{black}{~\cite{PierreAuger:2014ucz,ArteagaVelazquez:2023fda}.}

To this end, this paper details the first results of utilizing the world-class silicon imaging detector Subaru Hyper Suprime-Cam to observe individual tracks of EAS particles with unprecedented statistics and precision. In doing so, we develop a new method of observing EASs. 

\section{Observing extensive air showers with Subaru Hyper Suprime-Cam}

Subaru Hyper Suprime-Cam (HSC)~\cite{10.1093/pasj/psx063} is the prime-focus wide field camera on Subaru Telescope. Located atop Maunakea in Hawaii, it is the largest silicon imaging detector above an altitude of 4200\,m. Subaru HSC typically observes distant stars, galaxies and other interstellar objects in the optical and infrared. \textcolor{black}{This is achieved by employing 112 Charge Coupled Devices (CCDs) each with dimensions of 4176$\times$2048 pixels. Of these 112 CCDs, 104 are used for scientific observation and the remaining 8 are for calibration.
The total area of the CCDs which perform observations is $0.196$\,m$^2$.}

\begin{figure}
  \centering
  \includegraphics[width=0.99\linewidth]{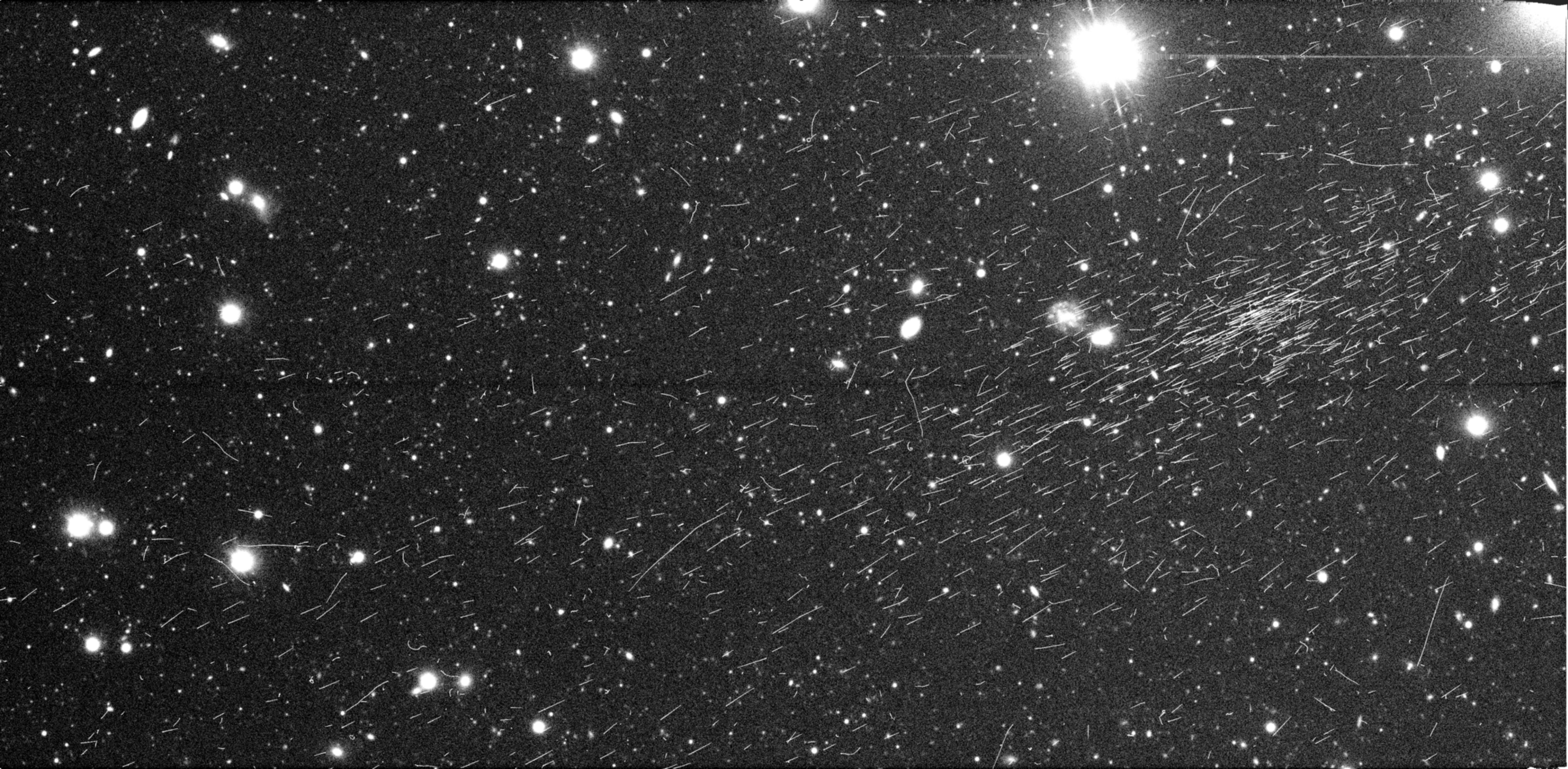}
  \caption{An example of a cosmic-ray extensive air shower recorded by a CCD of Subaru HSC. Dimensions : 4176 $\times$ 2048 pixels = 62.6\,mm $\times$ 30.7\,mm. The high density of aligned tracks, indicative of an EAS, is clearly visible in the upper right portion of the image.
  }
  \label{fig:HSC}
\end{figure}

\textcolor{black}{Standard operation of Subaru HSC involves taking images or ``exposures" of targets of interest throughout a night. Typical exposure times for individual images are 150 or 200 seconds. During an exposure charged particles from the atmosphere may penetrate into the depletion layer of the CCDs. These particles leave long, thin ``tracks" on the final image where the length of a track indicates the particle's angle of entry into the CCD. Normally, these tracks are sparse and randomly directed, and so are treated as unwanted noise which is removed. However, during the standard analysis of the Subaru HSC data, we encountered several images which displayed a far larger number of tracks than usual. Moreover, within each of these images, the majority of the tracks were aligned in a similar direction and were of similar lengths, indicating that the particles were travelling with similar trajectories. In turn we concluded that, for each of these images, the bulk of the detected particles had originated from a single, high-energy cosmic ray induced, EAS. We wish to emphasize the fact that the observation of these EASs was fortuitous and simply occurred during the normal operation of Subaru HSC. An example of one such image is shown in Figure 1.}


\textcolor{black}{Figure 1 contains several} particle tracks not aligned with the general direction of the shower. These may be deflected particles from the same shower, or randomly directed particles originating from the constant background of low-energy cosmic ray showers. Understanding the background rate of these randomly directed particles is important as, \textcolor{black}{based on how we initially found EAS events in our data}, we expect the number of particle tracks in an image where an EAS was present \textcolor{black}{to} be significantly greater than the background rate. Thus, by comparing the expected number of ``background tracks" in an image to the measured number, we can determine which exposures have observed an EAS. To achieve this, the following section constructs a model to describe the number of background tracks expected to be observed by Subaru HSC in different operating conditions.

\textcolor{black}{Before continuing, it is important to note that with only one silicon imaging detector, as is the case with our current data, it is not possible to determine the energy, arrival direction or mass composition information of the original cosmic rays. However, with additional detectors spaced in a similar fashion to a regular SD array, these properties could in principle be reconstructed. Additionally, we expect that the detailed measurements of individual secondary particles by silicon imaging detectors could be used in conjunction with standard SD arrays as an effective method for studying EASs in detail. Both of these points are discussed further in Section \ref{sec:discussion}.}

\section{Modeling the background rate of secondary particles}

\textcolor{black}{To effectively search for EASs} among Subaru HSC data, it is necessary to formulate a model for the number of randomly directed, background electromagnetic and muonic particles expected to leave tracks on a Subaru HSC image, given various data-taking conditions.
We label this value $N_{\rm{model}}$. The various contributions to $N_{\rm{model}}$ are described below, where the telescope elevation angle, telescope azimuth angle and sky brightness level are given by $\psi_{\rm{tel}}$,  $\phi_{\rm{tel}}$, and $B_{\rm{sky}}$ respectively. Zero degrees azimuth was defined to be south, with azimuth increasing clockwise.

\subsubsection*{Effective area}
We expect the number of detected tracks to be proportional to the effective detection area of the CCDs. This area itself is proportional to the absolute value of the cosine of angular separation between the telescope pointing direction and incoming cosmic ray direction.
Given the cosmic ray azimuth angle, $\phi_{\rm{CR}}$, and cosmic ray zenith angle, $\theta_{\rm{CR}}$, the cosine of angular separation is given by:
\begin{equation}
\cos \alpha = \cos \theta_{\rm{CR}} \sin \psi_{\rm{tel}} + \sin \theta_{\rm{CR}} \cos \psi_{\rm{tel}} \cos(\phi_{\rm{tel}}-\phi_{\rm{CR}})
\end{equation}
where $\alpha$ is angular separation.

Assuming that the zenith angle dependency of the cosmic ray flux at Earth is $\cos^{2} \theta_{\rm{CR}}$, the expected cosmic ray flux at telescope elevation $\psi_{\rm{tel}}$ is proportional to the solid angle integral

\begin{equation}
\int_{0}^{2\pi} \int_{0}^{\pi/2} \cos^2 \theta_{\rm{CR}} ~ | \cos \alpha | ~ \sin \theta_{\rm{CR}} ~ d\theta_{\rm{CR}} ~ d\phi_{\rm{CR}} = \frac{\pi}{4}(1+\sin^2\psi_{\rm{tel}}).
\end{equation}
Therefore, the term $(1+\sin^2 \psi_{\rm{tel}})/2$ is included in our model, where the normalization factor of $1/2$ has been chosen so that the term becomes unity at $\psi_{\textrm{tel}} = 90^{\circ}$ (west).

\subsubsection*{East-west effect}
The east-west effect is a suppression of low-energy positive cosmic rays arriving from the east due to the Earth's magnetic field~\cite{PhysRev.41.690,PhysRev.43.307,PhysRev.48.287}. With our definition of azimuth, this effect can be modelled as proportional to the term $\sin(\phi_{\rm{tel}})$.

\subsubsection*{Night sky background}
We define the night sky background, $B_{\rm{sky}}$, to be the median number of ADC counts across the CCD array over the period of an exposure. \textcolor{black}{This value changes with the background photon flux.} Empirically, we have observed that higher values of $B_{\rm{sky}}$
generally result in a larger number of tracks being detected. These additional tracks are believed to be ``false" triggers of the track detection algorithm, not actual secondary particles. Since the exact dependence of $B_{\rm{sky}}$ on the number of detected tracks is unknown, we assume a basic linear relationship in our model.\\

Adding the above effects, the final parameterisation of $N_{\rm{model}}$ is
\begin{equation}
\label{eq-model}
 N_{\rm{model}} = c_0 + c_1 (1+\sin^2(\psi_{\rm{tel}}))/2 + c_2\sin(\phi_{\rm{tel}}) + c_3 B_{\rm{sky}}
\end{equation}
where $c_0$, $c_1$, $c_2$ and $c_3$ are fitted parameters. Here, $c_0$ represents an offset to account for the different responses of the various filters used.

To find these parameters, we fit the above function to data collected by Subaru HSC between March 2014 and January 2020.
Subaru labels one measurement of the sky with a specific filter and exposure time as a ``visit". \textcolor{black}{For this study, visits with an exposure time of either 150 or 200\,seconds were used.} 
Table 1 below summarises the data showing, for each filter, the measurement period, number of visits in each period, exposure time for a single visit, and the corresponding total exposure time (subtotal). Across all filters, this gave approximately 875 hours of data.

\begin{table}
\begin{tabular}{lrrrrr}
\hline
Filter  &Start          &End            &\# of visits   &Exposure [s] & Subtotal time [s]\\
\hline
\hline
HSC-g   &Sep. 2014      &Jan. 2020      &2996   &150  &449400\\
HSC-r   &Sep. 2014      &Jul. 2016      &1079   &150  &161850\\
HSC-r2  &Aug. 2016      &Jan. 2020      &1737   &150  &260550\\
HSC-i   &Mar. 2014      &Nov. 2015      & 786   &200  &157200\\
HSC-i2  &Feb. 2016      &Jan. 2020      &3091   &200  &618200\\
HSC-z   &Sep. 2014      &Jan. 2020      &3887   &200  &777400\\
HSC-y   &Mar. 2014      &Jan. 2020      &3632   &200  &726400\\
\hline
Total time & & & & &3151000\\
\hline
\end{tabular}
\caption{Summary of the Subaru HSC data used to search for cosmic-ray extensive air showers.\label{table-data-1}}
\end{table}

For each visit, \textcolor{black}{dedicated} software was used to count the number of tracks in the image by searching for different morphologies such as a long-straight,  worm-like or wedge-like tracks. 
\textcolor{black}{The tracks were identified based on a series of conditions, primarily considering the pixel intensity relative to the local background and gradient checks against the point spread functions of the astronomical objects in the image. The procedure is described in depth in Section 4.4 of \cite{10.1093/pasj/psx080}. Pixels suspected to be contaminated by secondary particles were then grouped into contiguous events and further scrutinized in iterative passes.}

\begin{figure}
    \centering
    \includegraphics[width=0.8\linewidth]{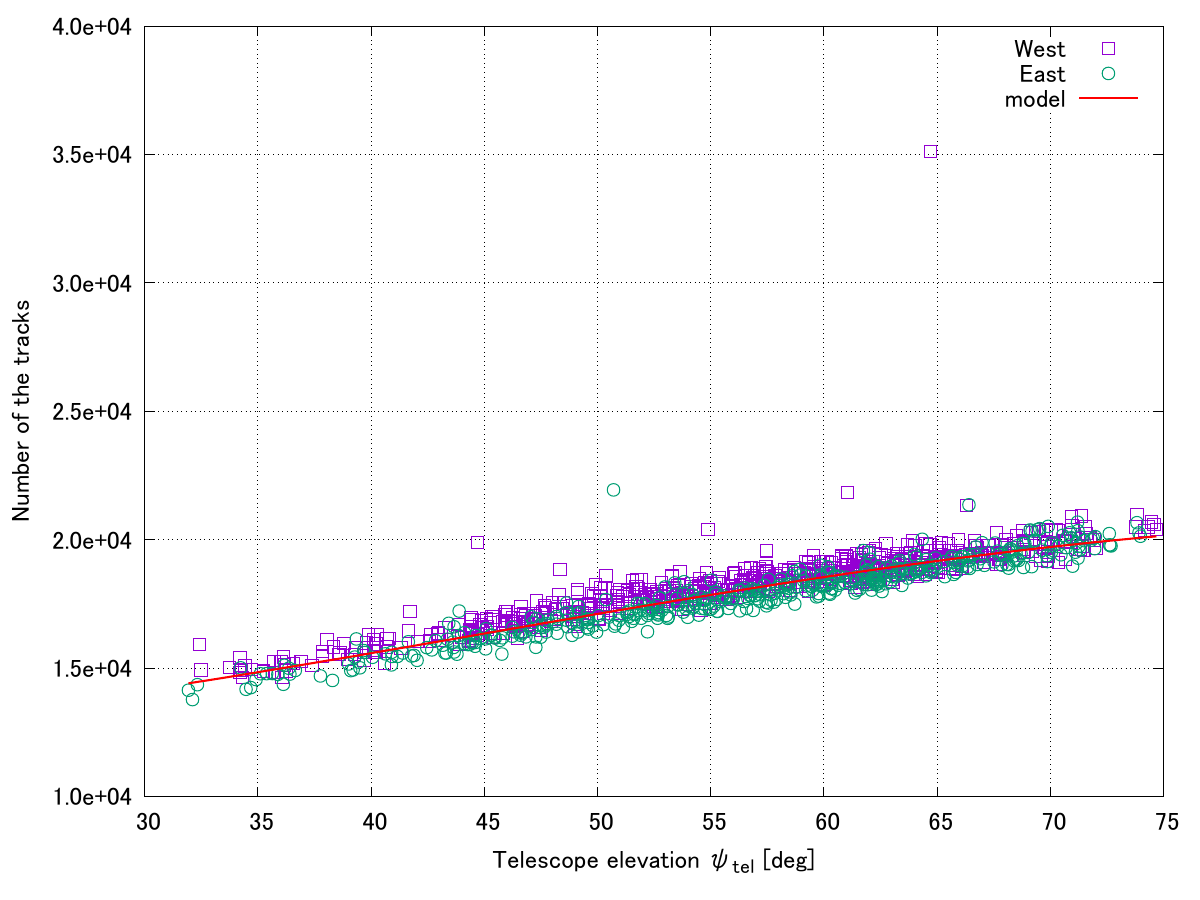}
    \caption{\textcolor{black}{A simplified example of fitting equation \ref{eq-model} to track numbers from different visits. The horizontal axis shows the telescope elevation angle. The vertical axis shows the number of detected tracks in a visit. The purple squares and green circles show data that was taken with the telescope pointing west and east respectively. There are several points above the main group of events. These points are likely to be EASs and are removed when applying the 2$\sigma$ threshold (see text). The red line shows the model fit results, with $\phi_{\rm{tel}} = 0 $ and $B_{\rm{sky}} = 1000$}.}
    \label{fig:example}
\end{figure}

\textcolor{black}{Once the number of tracks for each visit had been calculated}, the telescope's elevation and azimuthal angles, together with the night sky background levels, were extracted. \textcolor{black}{Equation \ref{eq-model} was then fit to the data from each filter using the following method. Initially, the model was fit to all the data, including events containing EASs, using a $\chi^2$ fit. As the objective was to model the number of \textit{background tracks} we then removed data points with a number of tracks different from the model prediction by $2\sigma$, where $\sigma$ was the standard deviation of the mean squared errors between the measured data and fitted model. Equation \ref{eq-model} was then fit again to the remaining data 
and the best fit parameters recorded.
} 
\textcolor{black}{We present an example of data and model fitting in Figure \ref{fig:example}.} 
The parameterisation results for all filters are shown in Table \ref{table-data-2}\ together with $\sigma$.

\begin{table}[ht]
\begin{tabular}{lcrrrrr}

\hline
Filter  &Material & $c_0$       &$c_1$  &$c_2$ &$c_3$   & $\sigma$\\
\hline
\hline
HSC-g   &silica& 1587$\pm$61  & 17939$\pm$69  &  235$\pm$8  &  0.185$\pm$0.027 &  329\\
HSC-r   &B270  & 3006$\pm$134 & 17600$\pm$146 &  289$\pm$14 &  0.141$\pm$0.027 &  341\\
HSC-r2  &silica& 1012$\pm$84  & 18437$\pm$96  &  169$\pm$11 &  0.208$\pm$0.014 &  336\\
HSC-i   &B270  & 8648$\pm$287 & 17819$\pm$285 &  249$\pm$28 &  0.271$\pm$0.023 &  506\\
HSC-i2  &B270  & 6480$\pm$105 & 19608$\pm$109 &  272$\pm$12 &  0.485$\pm$0.009 &  512\\
HSC-z   &silica& 2865$\pm$88  & 21749$\pm$92  &  272$\pm$10 &  0.494$\pm$0.009 &  448\\
HSC-y   &silica& 6409$\pm$126 & 18614$\pm$133 &  261$\pm$14 &  0.375$\pm$0.005 &  652\\
\hline
\end{tabular}

\caption{Results for the parameterisation of $N_{\rm{model}}$ for each filter.
\label{table-data-2}}
\end{table}

The dependency on elevation angle \textcolor{black}{($c_1$)} is roughly the same in every filter. Additionally, the azimuthal variation \textcolor{black}{($c_2$)} shows that the number of cosmic ray tracks from the west is greater than that from the east, as expected from the east-west effect.
The values for $c_3$ are comparatively small as typical values for $B_{\rm{sky}}$ are on the order of 1000 counts.
Finally, we note the large variation in $c_0$ across the different filters. In particular we see that, despite observing the same optical band, HSC-r has a significantly larger background level than HSC-r2. This may be because B270 contains potassium, a natural radio isotope.

\section{\textcolor{black}{EAS search results}}
\label{sec:result}
By using the model defined in the previous section, we search for visits in the data-set with a significant excess of tracks compared to our model prediction. \textcolor{black}{For demonstration purposes}, we have defined a significant excess to be more than 20$\sigma$ of each filter. Table~\ref{table-data-3} shows the significant visits. 
The telescope elevation angles for these visits, primarily between 50$^\circ$ -- 70$^\circ$, is simply a consequence of those elevations being the most commonly observed.

\begin{figure}
    \centering
    \includegraphics[width=0.49\linewidth]{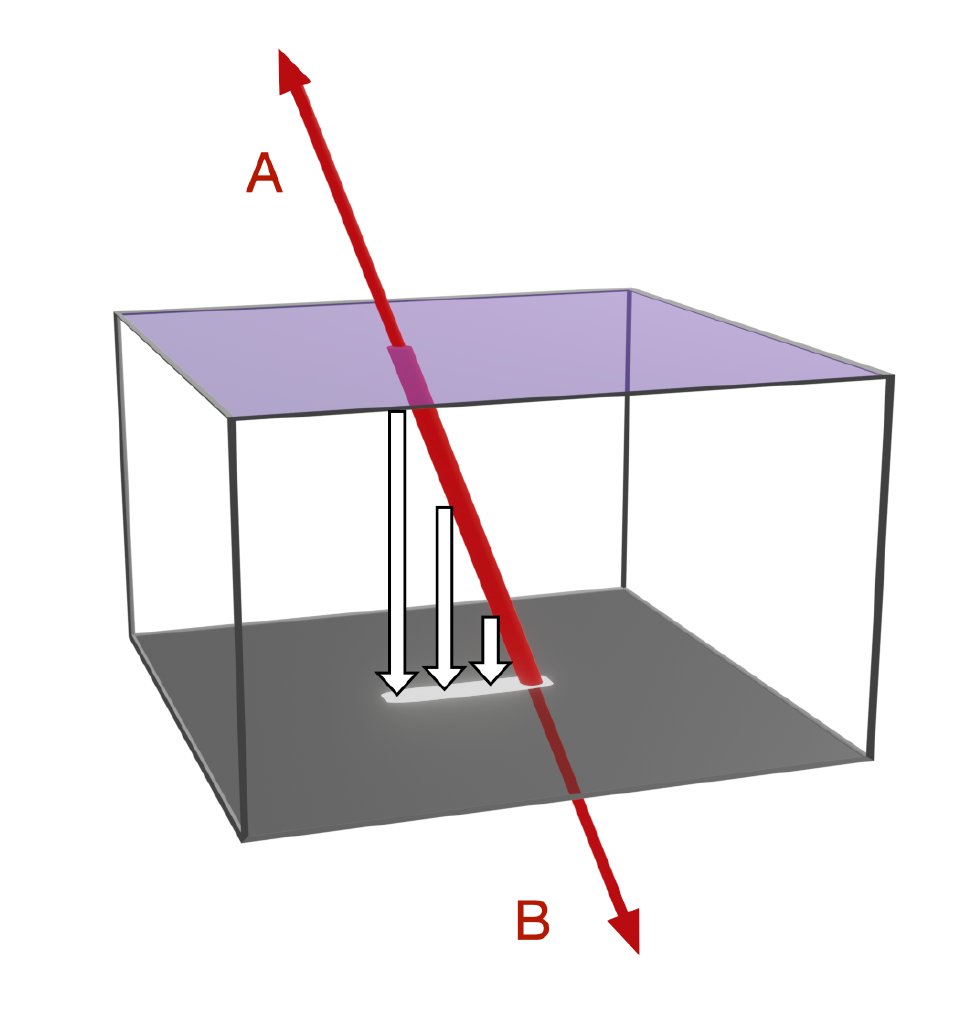}
    \includegraphics[width=0.49\linewidth]{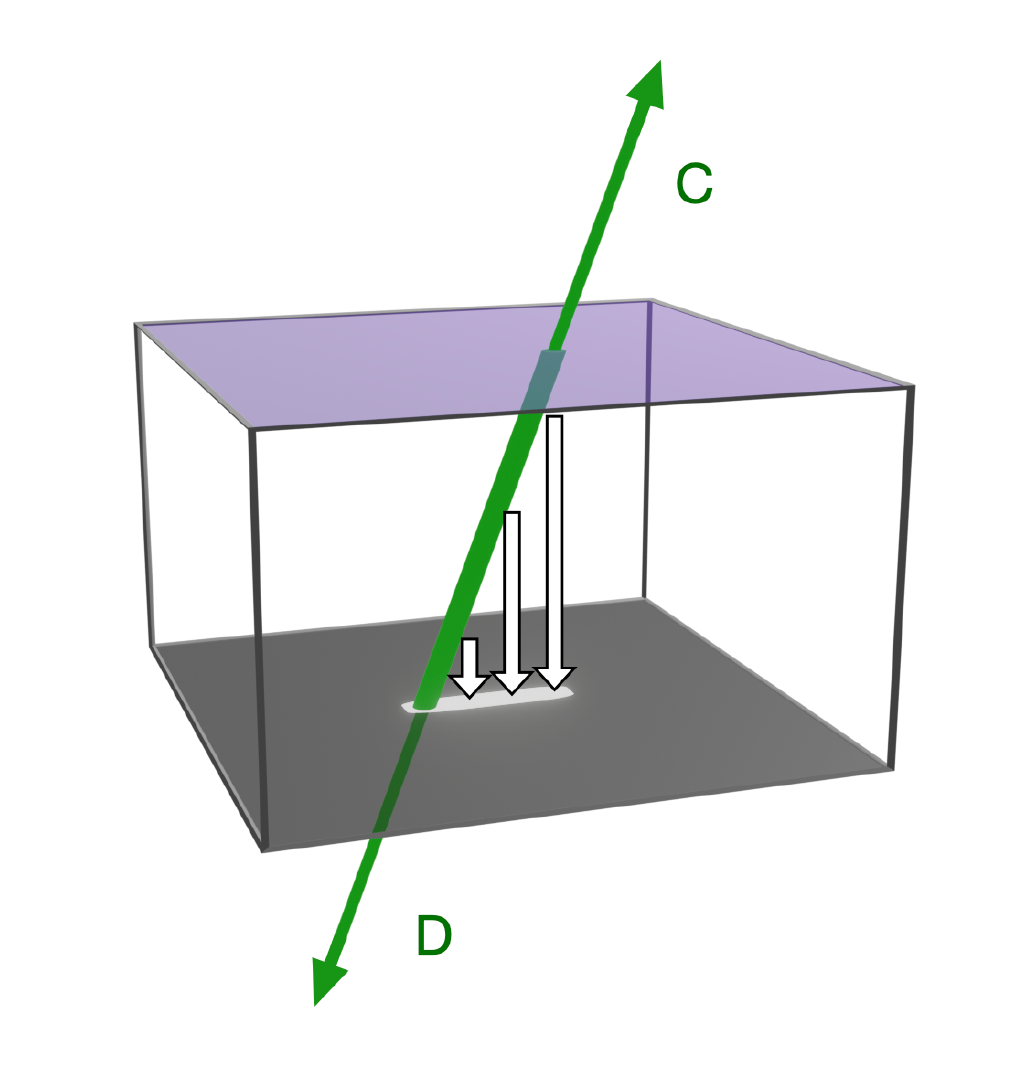}
    \caption{A schematic view of the four possible incident directions of a charged particle penetrating the depletion layer of a CCD. \textcolor{black}{The open arrows indicate the direction of charge transfer inside the depletion layer of the CCD. All four incident directions result in the same track being recorded. The up-going directions (A) and (C) are unlikely for cosmic rays (and hence their associated secondary particles).}}
    \label{fig:dir-deg}
\end{figure}

To determine whether these visits actually observed an EAS, \textcolor{black}{we checked for clustering in the particle arrival directions.
Unfortunately, the true 3D trajectories of each particle are impossible to determine with the CCD image alone. This is because there is no method for determining which end of a track corresponds to the top face/bottom face of the CCD. Furthermore, we cannot tell which end corresponds to the particle entering/exiting the CCD. In other words, for a single track there are four possible trajectories, each of which is shown in Figure \ref{fig:dir-deg}. If up-going particles are excluded (A and C in Figure \ref{fig:dir-deg}), then each measured track has two possible directions (B and D in Figure \ref{fig:dir-deg}). These directions differ by 180$^\circ$ in azimuth in the frame of the camera and have similar probabilities.} 
\textcolor{black}{For the sake of demonstration, we choose the trajectory with the smallest zenith angle. Regardless of whether this is the true direction, the observation of clustering will be evidence of an EAS.} 
Choosing one direction in this way limits the effective detection area of the telescope to $\pi$\,sr.

\begin{table*}[t]
\begin{center}
\begin{tabular}{ccccrrrrrr}

\hline
Visit & Date &      UT &       Filter &  $\psi_{\rm{tel}}$ &     $\phi_{\rm{tel}}$ &   $B_{\rm{sky}}$ &      $N_{\rm{track}}$&    $N_{\rm{model}}$ &      $N_{\rm{excess}}$\\

\hline
\hline
034298 &2015-07-14 &09:17:32 & g   & 54.2 &  138.9 &    413 &   24745 &   16685 &    8060\\
034480 &2015-07-14 &13:28:20 & g   & 72.7 &    1.4 &    423 &   28276 &   18760 &    9516\\
034814 &2015-07-15 &14:47:34 & r   & 64.7 &   45.5 &   1559 &   35124 &   19424 &   15700\\
039340 &2015-10-06 &14:08:04 & y   & 43.3 &   63.4 &   6209 &   36354 &   22328 &   14026\\
069450 &2016-04-15 &10:55:54 & y   & 55.5 &   54.3 &   7843 &   38513 &   24810 &   13703\\
073808 &2016-06-11 &09:42:43 & i2  & 59.0 &   57.1 &   2880 &   99476 &   25113 &   74363\\
146672 &2018-04-22 &09:42:33 & r2  & 63.2 &   50.7 &   2845 &   32509 &   18302 &   14207\\
161642 &2019-01-07 &15:31:21 & g   & 57.4 &   55.0 &    358 &   40683 &   17182 &   23500\\
162680 &2019-01-11 &05:20:15 & z   & 66.0 &   37.0 &   1874 &   33089 &   23908 &    9181\\
163754 &2019-02-02 &15:35:37 & g   & 51.7 &   68.7 &    367 &   23441 &   16365 &    7076\\
190348 &2019-11-01 &10:00:11 & g   & 60.7 &   61.1 &    418 &   26593 &   17657 &    8936\\
202364 &2020-01-03 &12:39:19 & g   & 51.1 &  -67.2 &    396 &   24279 &   15839 &    8440\\
203690 &2020-01-20 &14:27:54 & r2  & 67.3 &  -48.5 &   1017 &   27180 &   18166 &    9014\\
\hline
\end{tabular}
\caption{Event information of the possible extensive air showers detected by Subaru HSC.\label{table-data-3} ``HSC-" has been removed from the filter names for aesthetic purposes.}
\end{center}
\end{table*}

\begin{figure}
  \begin{center}
    \includegraphics[keepaspectratio,scale=0.65]{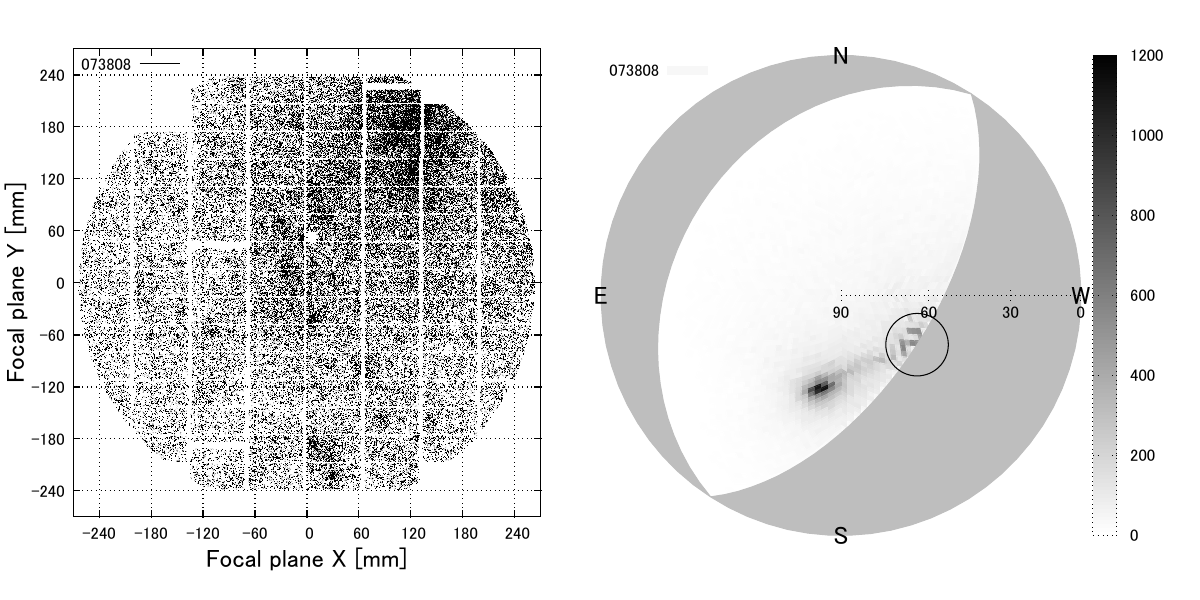}
    \includegraphics[keepaspectratio,scale=0.65]{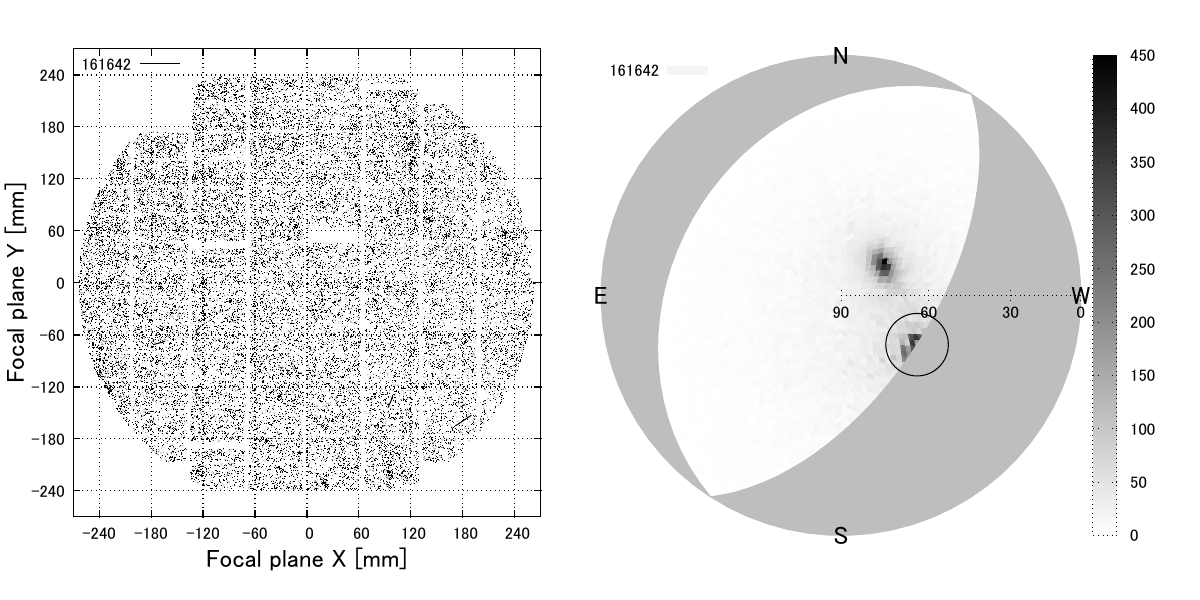}
    \caption{Arrival direction analysis for the largest and second largest excess visits of Subaru HSC. (Left) Positions of the secondary particle tracks in the CCDs. (Right) The traced back directions of the secondary particles in horizontal coordinates. The large open circles indicate the pointing direction of the telescope.}
    \label{muon-figure-17}
  \end{center}
\end{figure}

For each visit, the directions of the detected tracks were traced back to their position on the sky and plotted as 2D histograms. Two example histograms for visits 073808 and 161642, together with the positions of the raw tracks on the CCDs, are shown in Figure \ref{muon-figure-17}. The sky-map histograms have been divided into 12288 bins, with each bin equivalent to $\sim0.001$\,sr. A clustering of particle arrival directions is clearly visible for both visits, indicating a common source for these particles i.e. a primary cosmic ray. All significant visits show some level of clustering. The open black circles in Figure \ref{muon-figure-17} represent the telescope pointing direction, with these regions also containing a large number of entries.
\textcolor{black}{This is due to both the increased likelihood of detecting particles coming from the telescope pointing detection and} an artifact of the analysis procedure which causes the track detection algorithm to misinterpret stars, tracks which cross each other, and coincidental bundles of high signal pixels as short particle tracks.

\section{Discussion and future objectives}
\label{sec:discussion}
\textcolor{black}{For a single detector, the clustering observed in our results is encouraging and is evidence of having observed EASs. With additional silicon imaging detectors, the arrival direction, energy and mass composition of the primary cosmic ray may be able to be reconstructed. The particle density at each detector could be directly calculated based on the number of tracks and related to a primary energy through a lateral distribution fit, whilst the clustering of the secondary particle trajectories from \textit{multiple} detectors could lead to a reconstruction of the arrival direction. Assuming we are able to distinguish electrons/positrons from muons, which could be accomplished by separating curved (electron/positron) and straight (muon) tracks, mass composition information on an event-by-event basis may also be accessible through calculating the electromagnetic-muonic component ratio of the EAS.}

\textcolor{black}{Ultimately, we envision using silicon imaging detectors like Subaru HSC in conjunction with a SD array for highly detailed studies of EASs. 
With such a setup, the silicon imaging detector's ability to study EAS particle interactions in detail and determine event-by-event mass composition information could be combined with the precise knowledge of a showers energy and arrival detection from the SD array.} We intend to test this concept by installing plastic scintillators inside the Subaru HSC building. This will allow us to look for coincident events between the two detectors.



Moving forward, we intend to cross-check the results in Table \ref{table-data-3} with other telescopes at Maunakea Observatory, looking for time-coincident excesses evident of an EAS. If present, the particle densities at the different locations could be compared, similar to a SD array \textcolor{black}{as described above}, leading to a reconstruction of the primary energy.
Possible future interdisciplinary applications include using detailed measurements of the vertex of muon decay inside the CCDs for searches of lepton flavor violating interactions such as $\mu^{+} \rightarrow e^{+} + e^{-} + e^{+}$, similar to the Mu3e experiment~\cite{Mu3e:2020gyw}, and exotic signal searches for super heavy dark matter~\cite{PhysRevD.59.023501}.


\section{Conclusion}
We have reported on the first results of detecting cosmic-ray extensive air showers with Subaru Hyper Suprime-Cam. Visits with a significant excess of secondary particle tracks in the CCDs were found and analysed for clustering of particle arrival directions. All visits displayed such clustering, indicating a single high-energy cosmic ray origin for these particles. Additional \textcolor{black}{measurements} using a SD array and collaboration with other observatories will be critical for \textcolor{black}{taking full advantage of the unique properties of silicon imaging detectors as a tool for measuring extensive air showers.}


\bmhead{Acknowledgments}
This work was supported by JSPS KAKENHI Grant Number 20H00181, 20H05856, 22K21349, JP20H05852.
This work was supported by JST, the establishment of university fellowships towards the creation of science technology innovation, Grant Number JPMJFS2138. 
The authors appreciate Masanori Iye for fruitful suggestions of data analysis,
and thank Masato Yamanaka, Kohta Murase, Atsushi Naruko, Nagisa Hiroshima for valuable discussions regarding possible applications.

\bibliography{sn-article}%

\end{document}